# The Evolution of Agentic AI in Cybersecurity: From Single LLM Reasoners to Multi-Agent Systems and Autonomous Pipelines

Vaishali Vinay
vpapneja@microsoft.com
Microsoft Security Research
Redmond, Washington, USA

*Abstract*—Cybersecurity has become one of the earliest adopters of agentic AI, as security operations centers increasingly rely on multi-step reasoning, tool-driven analysis, and rapid decision-making under pressure. While individual large language models can summarize alerts or interpret unstructured reports, they fall short in real SOC environments that require grounded data access, reproducibility, and accountable workflows. In response, the field has seen a rapid architectural evolution from single-model helpers toward tool-augmented agents, distributed multi-agent systems, schema-bound tool ecosystems, and early explorations of semi-autonomous investigative pipelines.

This survey presents a five-generation taxonomy of agentic AI in cybersecurity. It traces how capabilities and risks change as systems advance from text-only LLM reasoners to multi-agent collaboration frameworks and constrained-autonomy pipelines. We compare these generations across core dimensions - reasoning depth, tool use, memory, reproducibility, and safety. In addition, we also synthesize emerging benchmarks used to evaluate cyber-oriented agents. Finally, we outline the unresolved challenges that accompany this evolution, such as response validation, tool-use correctness, multi-agent coordination, long-horizon reasoning, and safeguards for high-impact actions. Collectively, this work provides a structured perspective on how agentic AI is taking shape within cybersecurity and what is required to ensure its safe and reliable deployment.

## I. INTRODUCTION

Cybersecurity has become one of the earliest and most aggressive adopters of agentic AI. In contrast to most traditional enterprise domains, security operations centers (SOCs) continuously monitor high-volume telemetry, investigate multiple parallel alerts, and come up with threat hypotheses under severe time pressures [1]. Contemporary security workflows often include several interdependent steps: triage, enrichment, threat intelligence lookup, hypothesis building, evidence correlation, escalation, and reporting, so they are indeed multi-stage reasoning problems at heart [2]. The high level of cognitive load on analysts, continued understaffing, and increasing threat complexity have inspired the search for AI agents to help decision-intensive activities by assisting or automating these throughout the SOC lifecycle. Large language models (LLMs) were among the first AI systems shown to have strong usability in cybersecurity, summarizing alerts, extracting indicators from unstructured text, and explaining malware reports, but static LLM-only systems still don't suffice in realistic defensive environments [3]. For instance, LLMs continue to experience hallucinations, producing plausible but factually incorrect results [4]. More importantly, they lack grounded interaction with operational data sources, as they cannot independently pull telemetry from Security Information and Event Management (SIEM), endpoint detection and response (EDR), and threat-intelligence platforms, and their reasoning is stateless, without persistent memory or audit logs, which leads to inconsistencies over lengthy investigations [5]. To deploy in a SOC context with regulatory/compliance requirements, auditability, and verifiability are critical, but many LLM-only frameworks do not offer them [6]. Furthermore, without native tool integration, they may not be helpful in executing multi-step workflows in live environments.

Study of this kind is timely, as the cybersecurity industry is rapidly accelerating to commercial levels of AI-powered copilots, automated SOC modules, and early-stage autonomous response workflows. Nevertheless, not all vendor and open-source platforms provide a sufficient description of their agentic system architectural underpinnings, and the academic literature is disjointed, with previous reviews having mainly concentrated on LLMs, on the other hand, as an abstract concept (applications, vulnerabilities, defenses) rather than the structured development of agentic AI systems that are intended for defense in cyberspace. As a result, researchers and practitioners do not come up with one common evolutionary perspective to perceive how cyber-agents evolved as agents, how they acquire new competencies, and what risk they pose.

To address this void, this survey is informed by two high-level research questions: (1) What do we know about the architectural evolution of cybersecurity agents throughout time? and (2) As evolution progresses, what capabilities and risks emerge? There are four significant fundamental contributions achieved from this. First, the research presents a five-generation taxonomy of cyber-focused, agentic AI, from early single-LLM reasoners to fully autonomous cybersecurity pipelines. Second, the study presents a cross-generational comparative analysis of capabilities, multi-step reasoning, tool-use, memory, reproducibility, and safety. Third, the study synthesizes evaluation tools to benchmark gaps in the cyber-agent domain. Fourth, we suggest structured research agendas to develop safe, reliable, and verifiable agentic AI systems for cybersecurity.

## II. SCOPE AND METHODOLOGY

This survey focuses in particular on agentic AI systems for cybersecurity operations, especially on workflows of the Security Operations Center (SOC), threat-intelligence processing, detection engineering, and incident-response (IR) automation. Here, agentic AI means LLM-based reasoning is combined with tool usage, memory, planning, or verification capabilities that facilitate multi-step action execution as opposed to static text generation. This scope accords with new studies suggesting that cyber-defense tasks increasingly involve multi-hop reasoning in communication with operational sources of data [7], [8]. The survey clearly excludes adjacent but unrelated concepts, such as classic machine-learning classification models (malware detection, spam filtering), LLM-centric red-team automation, synthetic phishing generation, or generic chat-assistant models not built to structured cyber workflows. There is a lot of prior work on these domains, which do not reflect the architectural features of agentic, tool-based systems. The temporal boundary covering included works extends between 2020 and 2025, the period of the advent of GPT-3-scale models, the maturity of AI copilots with a security focus, and the introduction of structured tool-calling and agent frameworks. This span represents the transition away from independent LLM reasoning towards orchestration, multi-agent, and tool-based architectures. The study focuses on architectural evolution rather than product comparison or vendor assessment. This approach is justified by the recent findings showing that cybersecurity environments impose distinct and high reliability, auditability, and reproducibility requirements because incorrect or unverifiable automatic actions are commonly found [9], [10], [11]. Cybersecurity is chosen as the target domain of analysis given that it is home to an API-rich operational suite, including SIEM, EDR, SOAR, sandboxing, and threat-intelligence platforms, and it naturally delivers agent behavior supported by tools. Furthermore, SOC environments are decision-based and time-critical, so they serve as a good way to assess the limitations and the potential of agentic AI systems to be able to achieve real-world operational challenges
.

## III. BACKGROUND

The proliferation of LLMs is changing the way AI is used in cybersecurity, especially for interpreting unstructured text, reasoning over semi-structured data, and supporting analysts with high-level assessments. According to recent studies, LLMs have already been examined for cyber threat detection, threat-intelligence analysis, malware investigation support, and higher-level security reasoning, no longer confined to traditional classification pipelines or rule-based systems [12], [13]. These works indicate that LLMs provide assistance in activities such as finding IOCs, mapping text descriptions to threat techniques, and summarizing lengthy technical reports, reducing analysts' workload in large volumes [14], [15]. Simultaneously, contemporary security operations center (SOC) workflows are fundamentally tool- and data-based, and investigations are often performed relying on cross-correlating telemetry from SIEM platforms, EDR tools, network sensors, and cloud logs. It is emphasized in surveys of these SIEM technologies that they centralize log collection, real-time event correlation, incident monitoring & compliance reporting, functioning as the analytical backbone of several SOCs [16]. That way, any AI intended to be used for realistic SOC tasks must communicate with these systems rather than work directly on static, stored text.

Nevertheless, standalone LLMs face constraints that prove to be important, particularly in security, and they have all been found to hallucinate (producing fluent but factually incorrect statements), and the literature indicates that a certain degree of hallucination is intrinsic to models that generalize from their training distributions [17], [18], [19]. For long or multi-step investigations, these behaviors can undermine trust, particularly where outputs are not driven by live telemetry or coupled with explicit trace-making of their reasoning. Recent state-of-the-art reviews in the area of cybersecurity argue that LLMs facilitate innovative defensive capabilities; however, their reliability, controllability, and auditability are still key open challenges to mission-critical deployment [20]. These factors cumulatively drive a move from "LLM-as-a-chatbot" to agentic AI, systems in which LLMs serve as planners or controllers, calling tools, managing intermediate state, and working within orchestrated workflows. Current literature on the topic search looks into the application areas and limitations of LLMs in cybersecurity, yet it is not enough to trace the architectural development in the respective agentic systems over generations [10], [21], [22]. This gap paves the way for the current work, which focuses directly on the transition from single-model helpers to multi-agent, tool-integrated, and ultimately autonomous cybersecurity pipelines.

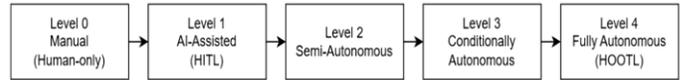

*Figure 1 Five levels of autonomy for AI-based SOCs [23]*

## IV. GENERATION 1- SINGLE LLM REASONERS

AI security tools have built upon standalone LLMs, for instance, GPT-3 and GPT-4, which work only in text space. These Gen-1 AI agents can perform natural language processing efficiently: they summarize long cyber reports in accessible language to summarize unstructured security warnings, help the analyst make sense of them, and respond to analyst queries with contextual reasoning and the like [7], [24]. Their best feature is rapid sense-making, as an LLM can parse a phishing email, extract IOCs, recognize social-engineering clues, and assess intent in a coded way similar to human instinctual code, at an impressive speed. In response, models can condense a long malware analysis into executive summaries or associate alert descriptions with MITRE ATT&CK techniques, thereby relieving cognitive burden for beleaguered SOC [7], [24]. In practical usage, it is the case that more commonly Gen-1 LLMs are being employed as assistive copilots among analysts. In the face of noisy alerts, the model can demystify probable root cause, map relevant TTPs, or regroup jargon-laden intelligence reports into plain language to inform briefings, but even with those beneficial capabilities, Gen-1 systems are still fundamentally restricted, and one of the reason is that they don't run external queries, can't fetch data, run code, or talk to security tools [7].

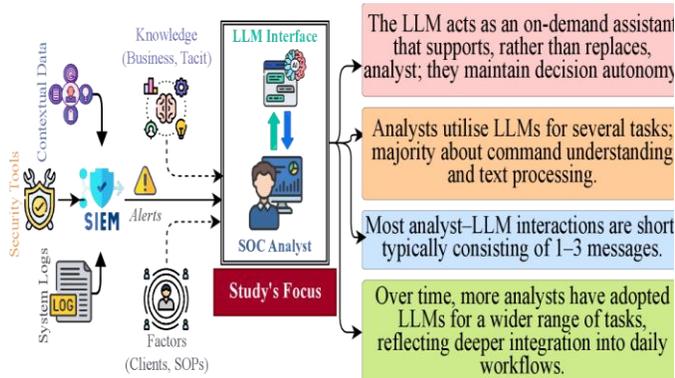

Figure 2 SOC Workflow [24]

Every response is "stateless," as the model has no persistent memory and cannot pass information from query to query if not given explicit information. These constraints of the architecture bring with them observed vulnerabilities. While models work in probability space and do not use external verification mechanisms, if the prompt is ambiguous or data is found missing, it hallucinates or misses crucial details [23], [25]. The output of an even relatively minor prompt change can be very different, and consequently, the reproducibility will be poor. Further, because it cannot carry out actual acts, whether by reviewing logs, accessing CVE registries, or parsing output from the sandbox, they all rely upon internal pattern recognition, rendering them prone to confident yet false conclusions. So, Gen-1 LLMs can indeed produce some powerful text processing, but they are powerless to self-validate the knowledge they have developed and to act. This instability restricts them to ineffective decision-making in a high-stakes context. They remain suitable for summarizing, explaining, and threat-intel digestion [7], but analysts need to consider their outputs in a more advisory approach than authoritative. Ultimately, Gen-1 provides a valuable but limited base: a strong linguistic intelligence coupled with structural weaknesses that stop reliable automation [23], [25].

## V. GENERATION 2 - TOOL-AUGMENTED AGENTS

Gen-2 systems were created to rectify the main shortcoming of Gen-1, which is the lack of action. These agents leverage the reasoning techniques from LLM as well as external tools to execute queries, run filters, access APIs, and generate structured output. Architectures including ReAct and Planner–Executor present a structured loop where the agent reasons over iterated steps, selects an action, executes a call to tools, observes results, and updates its internal plan [23]. This turns the LLM from a passive text generator into an enabler for validating information. Within cybersecurity workflows, Gen-2 agents provide significant enhancements. They have the ability to automate SIEM or data-lake queries by converting the natural language instructions into structured search expressions. Analysts can ask "Show events related to this suspicious IP," and the agent prepares and runs the correct query against the log store. Similarly, agents can supplement alerts by consulting WHOIS services, CVE databases, passive DNS sources, and threat-intelligence platforms for authoritative facts instead of hallucinating [23], [26], [27], [28]. This tool-enhanced pattern also allows IOC extraction and pivoting, and an LLM reads a threat report, isolates hashes or domains, and forwards them to an external API for supporting evidence. The change from pure language to action-enabled workflows greatly improves reliability.

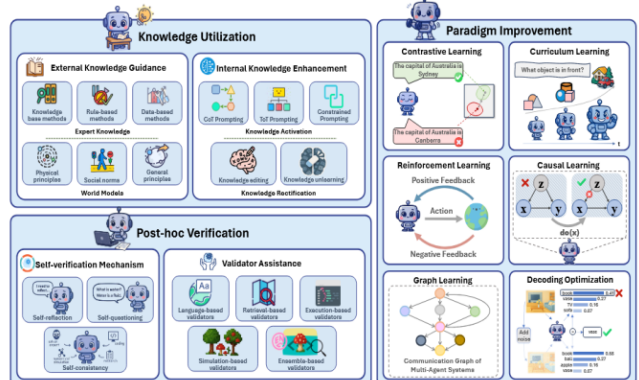

Figure 3 Hallucination mitigation [29]

By the LLM checks assumption, which is whether an indicator occurs or not, for instance, in logs, it can help to reduce unsupported claims. However, introducing multiple steps also brings new failure modes. Mistakes can propagate through the chain: an erroneous early hypothesis can cause the agent to generate a suboptimal query, which gives erroneous output, exacerbating the agent's error. These cascades are particularly dangerous given that agents frequently construct iterative rationales, which become more and more entrenched with the first error [29]. Tool-based execution also involves the fragility of API availability: if one service goes down or returns incorrect data, the entire logic pipeline can crack. Centralization is another architectural risk. Most Gen-2 systems involve one LLM instance as the planner and executor. If the model outputs a wrong plan or misinterprets the output from the tools, the pipeline is alone with no secondary layer to intervene. Consequently, the system is now one single failure point. A high-risk, high-stakes cybersecurity operation (e.g., incident response) cannot afford such a risk without human oversight. Despite these constraints, Gen-2 agents significantly expand SOC capabilities, including semi-automated alert triage, log investigation, and contextual enrichment [23]. They're excellent at being analyst amplifiers and not individual operators. Their strength comes in joining natural language and the real data, but they are vulnerable to error drift and ripple-level logic failures that only their persistent monitoring can alleviate [23], [29].

## VI. GENERATION 3 — MULTI-AGENT CYBER SYSTEMS

Gen-3 represents a structural move away from single agents to distributed role-based teams of collaborating agents. Rather than having one LLM plan, execute, and evaluate tasks, the systems now orchestrate a series of specialized agents sharing context, critiquing one another, and modularizing the incident-response pipeline [30], [31], [32]. This is similar to the structure in which real SOC teams are defined, composed of triage analysts, investigators, and reviewers. There are generally 3 core agents for a design pattern: a triage agent, an investigator agent, and a verifier/critic agent. The triage is where the context of incidents is determined, alert severity is classified, and the plan for investigation is developed. The investigator agent sorts through search information, parses logs, enriches IOCs, and links evidence between sources of data. Finally, the verifier actor tests whether conclusions can be deduced from evidence and detects hallucinations, inconsistencies, or otherwise significant and often missing data. This multi-level architecture is more robust and leads to more accountability. Prototypes of real-world research illustrate these principles. The CORTEX system employs a Behavior Analysis agent in its search engine interface to identify relevant workflows, Evidence Acquisition agents to execute SIEM or threat-feed queries to help in generating action plans, and lastly, a Reasoning agent to aggregate this information into an auditable triage decision [33]. IRCopilot is a third example that uses Planner, Generator, Reflector, and Analyst agents to simulate a full range of incident response processes and documentation [34]. Open-source ecosystems such as Microsoft AutoGen also make multiple-agents orchestration libraries available for developers to create similar modular workflows.

There are several advantages of the multi-agent paradigm. It allows the multi-step multi-task complexity (i.e., associating multiple log sources) to be passed between agents, leading to more reliable reasoning [33], [34]. Second, cross-verification among agents lowers the chance of unchallenged hallucinations. Third, multi-agent systems can maintain shared long-horizon context during an incident, which improves memory continuity compared with single agents, but Gen-3 has its very own class of failure modes. Agents may have competing hypotheses, resulting in diverging paths to investigate [33], [34]. Poorly constrained communication protocols may lead to infinite loops wherein agents keep saying they need clarification or another plan without converging. The CORTEX study highlights risks like "inter-agent misalignment" and "specification errors," or scenarios where agents view roles or instructions differently from the intended ones [33], and this may result in redundancy for actions, inconsistent results, or fluctuating behavior. In a similar vein, multi-agent systems exacerbate feedback loops: a mistake from one agent that receives its stamp of approval from another can proliferate fast across the system. These problems make rigorous coordination and termination rules extremely necessary. Without guardrails, multi-agent systems (and with them a lot of complex reasoning tasks) could lead to a far longer and less understandable result, compared to the output produced by one model alone. However, despite these problems, Gen-3 is still an avenue toward complete SOC workflows because it distributes the cognitive load and enables multi-tiered quality checks [33], [34].

VII. GENERATION 4 -MCP-BASED & STANDARDIZED TOOL ECOSYSTEMS

Agentic cybersecurity systems in the fourth generation are a significant development in terms of stability, predictability, and regulation, thanks to tool schemas that are standardized [35]. Whereas in earlier generations, agents would spontaneously make API calls or create ad hoc command structures, the Gen-4 systems are highly circumscribed. Agents interact with other enterprise security solutions like SIEM platforms, EDR agents, threat-intelligence systems, and SOAR pipelines, using a set of pre-defined schemas that dictate the formats in which each interaction is generated [35]. This schema is often translated into a JSON contract, function call structure, or a domain-specific interface that makes sure there is a predictable syntax and expected output that underpins every agent–to–tool exchange [23]. Frameworks have formalized these rules, such as the MCP, creating explicit parameters for input and output per tool that an LLM is allowed to call.

This method provides several essential advantages, like standardization increases reproducibility to begin with, and the tool call provided today, when the inputs are the same, will produce the same result tomorrow, because the agent should execute following the same schema, and it cannot fall into unstructured or ambiguous instructions. This is in stark contrast to the Gen-1 and Gen-2 systems, for which small prompt changes or LLM variability could alter system behavior [35]. Second, using a schema-dependent approach in interaction enhances safety, as agents cannot, for example, run arbitrary shell commands, change high-impact settings, or provide malformed requests, because each of these tools has a whitelisted set of callable functions. Each request must be typed, verified, and recorded, allowing for update, edit, and audit for version control and security. Agents generated via playbooks can produce standard CACAO-formatted workflows with standardized workflows instead of all other free text as well, allowing remediation instructions to be syntactically correct and machine-actionable [36]. Another unique advantage of Gen-4 buildings is in governance and compliance. All tool invocations are recorded with full transparency, including which schema/schemas, what parameters were retrieved, which data sources were used, and what results were returned, and this fits with enterprise audit requirements, incident-response documentation requirements, and regulatory controls [37], [38].

In a SOC setting, such detailed provenance helps enable an analyst to examine AI-assisted actions over time, compare outputs, and identify unauthorized deviations. The bounded execution environment also reduces incidental misuse; an LLM, for example, cannot disable a firewall rule, in the absence of these actions existing in itself as schema-controlled operations. That said, the Gen-4 systems are not without their limitations. Whilst schemas strongly dictate what action is done, they don't necessarily determine the semantic accuracy of the action, an LLM will misinterpret the evidence that is given to it at hand, so it will produce a syntactically correct but context-poor log query that ends up being incorrect, or maybe it will pick out the right tool to use to run the investigation, technically wrong [39], [40]. These semantic failures are still the most frequent because LLM-based reasoning still relies on a natural language probability space. Even when all tool calls correspond to the exact schema definitions, hallucinations might still arise in the agent's internal cognition that could guide the subsequent steps [23]. This gap provides a reason why MCP-based approaches still need human supervision. As indicated throughout the literature, LLM-produced security actions "require human verification of outputs," as hallucinated interpretation, inappropriate comparisons, or unfair conclusions still are not acceptable in active SOC environments [23]. Schemas do not implement correct analysis, but correct structure, and the model could correctly make a call to a SIEM API, but that means it should not only give incorrect responses, but also misjudge why the output was returned, it may misclassify severity, and/or suggest methods of remediation contrary to recommended best practice. Overall, Gen-4 architectures implement strict execution using a schema that will lead to increased reproducibility, auditability, and operational safety [36]. They give frameworks for enterprise adoption, but cannot automatically ensure analytical accuracy. Human validation and upper-level verification are still necessary to confirm that planned, well-formed actions are also contextually correct and secure.

## VIII. GENERATION 5 - AUTONOMOUS CYBERSECURITY PIPELINES

The fifth generation represents the most conceptually ambitious stage, aiming to support end-to-end SOC workflows with minimal human intervention [41], [42]. Rather than needing incremental nudges and operator-centered supervision, Gen-5 agents act based on high-level goals. In these, an analyst could issue a single order like "Investigate this breach," "Analyze this alert," or "Generate an incident report," and the agent establishes an entire investigative pipeline. These include extracting the appropriate alerts from SIEM platforms, correlating telemetry from EDR or network logs, enriching the indicators with threat-intelligence lookups, and aggregating data into an orderly product usable by humans. The construction of these systems is highly modular. A Gen-5 IR assistant could, for instance, autonomously pull raw alerts, map them to MITRE ATT&CK techniques, execute tailored log queries, conduct anomaly detection, and piece together correlated evidence from several data platforms and sources [43], [44]. This includes creating a situational summary of the research process, drafting a mitigation playbook, or recommending actions for containment. Systems like Microsoft Security Copilot or academic prototypes like IRCopilot exhibit elements of this idea, with discrete components for detection, investigation, containment, and remediation, each with clear roles [34]. In experimental results from multi-phase research systems, the most impressive enhancements in efficiency are observed; IRCopilot reports up to ~150% more tasks completed compared to a traditional single monolithic LLM, which emphasizes what is possible with semi-automated orchestration with constrained autonomy [34].

Gen-5 systems have several potential uses, encompassing both proactive and reactive security roles. Prototype Gen-5 autonomous IR assistants aim to automate portions of workflow from ingestion to final reporting with little human interaction, and other uses include automation/inspection detection logic, where agents autonomously adjust IDS/IPS signatures or SIEM correlation rules based on observed threat behavior [45]. Gen-5 systems may autonomously generate new detection content, validate hypotheses across multiple data environments, and even plan partial containment workflows when malicious behavior is confirmed, but the power of Gen-5 systems comes with equally grave risks. By having autonomy, agents can run directly on live infrastructure, even issuing high-priority commands. A simulated or incorrectly mapped threat scenario could cause an agent to block production servers, close functioning processes, change firewall rules, and shut down mission-critical services without human intervention. Unverified autonomy may unintentionally trigger disruptive actions such as extended outages, security holes, and a chain reaction of operational failures, underscoring the need for strict safeguards.

The dangers go well beyond operational disruption: as an autonomous agent that handles sensitive log data, users' IDs, and internal telemetry, the literature repeatedly warns against "data leakage" and privacy breaches when AI agents are operating unmonitored [23]. In addition, without human-in-the-loop, a faulty investigative chain that is anchored only in a single wrong assumption could spread unchecked throughout the entire system, resulting in erroneous remediation strategies. Thus, Gen-5 implementation needs to have strict security mechanisms applied. SOCs will need to implement read-only modes for default behaviors, require manual approval by others to change system state, and have layered verification of high-impact decisions [23]. Accountability requires continuous monitoring, reliable telemetry, and transparent decision-making across agents. Gen-5 systems, in practice, ought to augment analysts, like free assistants whose recommendations have to be verified by humans themselves before they can be rolled out.

## IX. CROSS-GENERATION CAPABILITY COMPARISON

At five generation levels, agentic AI capabilities deepen on a number of important fronts. Gen-1 LLMs offer only single-step reasoning and limited TTP mapping based largely on static recall rather than grounded analysis [25]. Gen-2 systems are capable of simple multi-step planning when applying tool calls and can automatically search log records and simple MITRE ATT&CK retrieval, but still lack long-term memory and reliable reproducibility [23]. Gen-3's multi-agent workflows greatly support task specialization, facilitating complex multi-phase IR pipelines in which agents perform triage, analysis, evidence correlation, or reporting [33], [34]. Common context amongst agents can likewise promote long-horizon memory. Gen-4 expands reproducibility even more through schema-bound API interactions to keep tool calls logged, typed, and replayable [36]. And those very rigid interfaces also make the program much safer by guarding against malformed or unauthorized operations. Gen-3 uses cross-agent verification, and Gen-4 uses schema validation, both of which offer stronger protections than those offered by the generations preceding them. Trade-offs remain, though: the earlier generations restrict but make things predictable, while later generations provide a greater degree of autonomy to the detriment of new failure scenarios (e.g., conflicting agent outputs, semantic misunderstanding, cascading errors, unintended automation) [23], [33]. Cumulatively, however, capability gains are accompanied by increasing requests for oversight, verification, and control.

*Table 1 Comparison*

| Capability Dimension | Gen-1 | Gen-2 | Gen-3 | Gen-4 | Gen-5 |
|---|---|---|---|---|---|
| **Reasoning Depth** | Single-step | Limited multi-step via tools | Multi-agent multi-step | Structured reasoning with schemas | Full autonomous pipelines |
| **TTP Mapping** | Static recall | Simple retrieval-based | Distributed mapping via agents | Schema-validated mapping | Autonomous contextual mapping |
| **Memory** | None | Short, prompt-bound | Shared long-horizon memory | Reproducible via logs | Persistent pipeline-level memory |
| **Reproducibility** | Low | Moderate | Higher via cross-agent checks | High via schemas | High but risk-sensitive |
| **Safety & Verification** | Weak | Error-prone | Critic/validator agents | Strong schema boundaries | Requires strict safeguards |

## X. EVALUATION LANDSCAPE AND RESEARCH CHALLENGES

There are various benchmarks out there for measuring AI agents, but the evaluation space is far from comprehensive and not always even. As a general-purpose assessment tool, such as AgentBench, we evaluate autonomous reasoning under a wide variety of simulated environments. DefenderBench is mainly used for cyber offense and defense types of tasks, like intrusion simulation, exploit generation, and malware analysis. In security-facing settings, CyBench provides CTF-style tests, and SecEval offers structured security knowledge assessments [46]. MITRE's ATLAS framework (v2) captures adversarial AI attack patterns to guide the targeting and exploitation of agentic systems [47], [48]. Even with these advances, there are profound capability inadequacies. Perhaps most importantly, there is no available benchmark that is capable of measuring end-to-end SOC workflows from alert detection to investigation, enrichment, correlation, and final reporting in actual service use cases. This makes it an essential omission as multi-step agentic behavior provides the basis for Gen-3, Gen-4, and Gen-5 systems. No less missing are metrics for multi-agent collaboration, from division of labor to conflict resolution or consensus formation. And there is no common standard for assessing "tool-use correctness," which is to say, if an agent's API calls were not only syntactically correct but also semantically suitable and operationally effective. Verifier and critic agents, key traits of Gen-3 and Gen-4, lack common effectiveness metrics. Most benchmark datasets, however, (despite their potential) do not come with the noise, ambiguity, and contradictory evidence found in real SOC environments, or other mixed evidence of multiple alternatives. As discussed in existing evaluations, a realistic multi-step SOC benchmark with strong coverage of planning, memory, tool chaining, and interplay of agents on a large scale is urgently needed in the field.

*Table 2 Benchmarking*

| Benchmark | Reasoning Tasks | Tool Use | Cyber-Specific Tasks | Multi-Agent Evaluation | Long-Horizon Tasks | Security Sensitivity | Notes |
|---|---|---|---|---|---|---|---|
| **AgentBench (2023–2024)** | ✓ | ◐ (limited) | ✗ | ✗ | ◐ | Medium | General-purpose; limited cyber depth [49] |
| **DefenderBench (2024–2025)** | ✓ | ✓ | ✓ | ✗ | ◐ | High | Strong cyber tasks; single-agent focus [46] |
| **CyberSOCEval (2025)** | ✓ | ◐ | ✓ | ✗ | ✗ | High | Malware + TI reasoning; no orchestration [50] |
| **CyBench / CyberBattleSim Tasks** | ✓ | ✓ | ✓ | ✗ | ◐ | High | CTF-style; offensive leaning [51] |
| **SecEval** | ✓ | ✗ | ✓ | ✗ | ✗ | Medium | Security Q&A; no tool use [52] |
| **AttackSeqBench** | ✓ | ✗ | ✓ | ✗ | ✗ | Medium | Adversarial AI + ATT&CK mapping [53] |
| **AutoPenBench / Red-Team Pentest Benchmarks** | ✓ | ✓ | ✓ | ✗ | ◐ | High | Offensive; exploit-focused [54] |

These shortcomings highlight deeper research challenges. An urgent need exists for verification and critical models to independently analyze the correctness of agent outputs, recognizing hallucinations, invalid correlations, or unjustified conclusions [23], [29]. Limits of safety are equally important here, because if Gen-5 systems will work autonomously, they should have reasonable restrictions or fail-safes to safeguard against destructive actions, misconfigurations, or data exposure [23]. To ensure cross-agent consistency of multi-agent systems, it has become challenging to perform research on consensus protocols and shared world-model alignment. Standard schemas, like CACAO for playbooks [36], should be modified to include SIEM, EDR, TIP, and SOAR tools with full semantic definitions. Other problems are long-horizon planning and memory retention, since existing models tend not to be able to apply well across multiple workflows. There are ethical and privacy concerns as well, as agentic systems commonly handle sensitive log records, internal telemetry, and individual user information, needing strict checks and balances to prevent bias, leakage, and maintain accountability [23].

## XI. Conclusion

By and large, this taxonomy of generations shows progress, from 1-step Gen-1 models to multi-agent collaboration and to autonomous Gen-5 pipelines, but also underscores the parallel rise in risk and sophistication. As emphasized across the literature, the future of AI's dependable security depends on robust standards, evaluation frameworks, and principled oversight to ensure deployment that is safe, reliable, and verifiable.